\documentclass{aa}
\usepackage{natbib}
\usepackage{xspace}
\usepackage{amssymb}
\usepackage{amsmath}
\usepackage{graphicx}
\def\cdrev#1{#1}
\def\keesrev#1{#1}
\def\aap{A\& A}

\def\araa{AnnRevA\& A}
\def\mnras{MNRAS}

\def\apj{ApJ}

\def\apjl{ApJL}

\def\gram{\hbox{g}}
\def\cm{\hbox{cm}}

\def\AU{\hbox{AU}}

\def\vset{\ensuremath{v_\mathrm{sett}}\xspace}

\def\cs{\ensuremath{c_\mathrm{s}}\xspace}
\def\siggas{\ensuremath{\sigma_\mathrm{g}}\xspace}
\def\sigcoag{\ensuremath{\sigma_\mathrm{c}}\xspace}

\def\rhogas{\ensuremath{\rho}\xspace}

\def\Hp{\ensuremath{H_\mathrm{p}}\xspace}

\def\Omegak{\ensuremath{\Omega_\mathrm{K}}\xspace}

\def\Sc{\ensuremath{\mathrm{Sc}}\xspace}
\def\St{\ensuremath{\mathrm{St}}\xspace}

\def\comma{\,,}
\def\fullstop{\,.}
\def\alphairr{\ensuremath{\alpha_\mathrm{irr}}\xspace}
\def\alphaturb{\ensuremath{\alpha_\mathrm{turb}}\xspace}

\begin{document}
\title{Dust coagulation in protoplanetary disks: a rapid depletion of small grains}
\titlerunning{Dust coagulation in protoplanetary disks}
\authorrunning{Dullemond, Dominik}
\author{C.P.~Dullemond \& C.~Dominik}
\institute{Max Planck Institut f\"ur Astrophysik, P.O.~Box 1317, D--85741 
Garching, Germany; e--mail: dullemon@mpa-garching.mpg.de\\
Sterrenkundig Instituut `Anton Pannekoek', Kruislaan 403,
  NL-1098 SJ Amsterdam, The Netherlands; e--mail: dominik@science.uva.nl}
\date{\today}

\abstract{We model the process of dust coagulation in protoplanetary disks
and calculate how it affects their observational appearance. Our model
involves the detailed solution of the coagulation equation at every location
in the disk. At regular time intervals we feed the resulting 3-D dust
distribution functions into a continuum radiative transfer code to obtain
spectral energy distributions. We find that, even if only the very basic --
and well understood -- coagulation mechanisms are included, the process of
grain growth is much too quick to be consistent with infrared observations
of T Tauri disks. Small grains are removed so efficiently that, long before
the disk reaches an age of $10^6$ years typical of T Tauri stars, the SED
shows only very weak infrared excess. This is inconsistent with observed
SEDs of most classical T Tauri stars. Small grains somehow need to be
replenished, for instance by aggregate fragmentation through high-speed
collisions. A very simplified calculation shows that when aggregate
fragmentation is included, a quasi-stationary grain size distribution is
obtained in which growth and fragmentation are in equilibrium. This
quasi-stationary state may last $10^6$ years or even longer, dependent on
the circumstances in the disk, and may bring the time scales into the right
regime. If this is indeed the case, or if other processes are responsible
for the replenishment of small grains, then the typical grain sizes inferred
from infrared spectral features of T Tauri disks do not necessarily reflect
the age of the system (small grains $\rightarrow$ young, larger grains
$\rightarrow$ older), as is often proposed. Indeed, there is evidence
reported in the literature that the typical inferred grain sizes do not
correlate with the age of the star.  Instead, it is more likely that the
typical grain sizes found in T Tauri star (and Herbig Ae/Be star and Brown
Dwarf) disks reflect the state of the disk in some more complicated way,
e.g.~the strength of the turbulence, the amount of dust mass transformed
into planetesimals, the amount of gas lost via evaporation etc. A simple
evolutionary scenario in which grains slowly grow from pristine $0.1\mu$m
grains to larger grains over a period of a few Myr is most likely
incorrect.}

\maketitle

\begin{keywords}
accretion, accretion disks -- circumstellar matter 
-- stars: formation, pre-main-sequence -- infrared: stars 
\end{keywords}

\section{Introduction}
The coagulation of dust grains in protoplanetary disks is believed to
be the first stage of planet formation. The dust grains, inherited
from the interstellar medium, are initially particles of submicron
size. But as a result of the high densities in the disk, and due to
processes such as Brownian motion, settling and turbulence, the dust
grains quickly get in contact with each other, and generally stick and
form aggregates of ever increasing size. It is believed that the
continuing growth of such grains/aggregates eventually leads to the
formation of planetesimals, which, through gravitational interaction,
coalesce to form the rocky cores of planets.
The subsequent accretion of gas onto these cores, if they are
massive enough, then leads to the formation of gaseous giant planets.
This is known as the core accretion model for giant planet formation
(Lissauer \citeyear{lissauer:1993}).

A significant body of theoretical research has so far mostly focussed
on our own solar system, trying to explain the structure of the system
as well as the relevant time scales known from geological and
meteoritic research.  Theoretical models are constrained for example
by the fact that the core of Jupiter must have formed fast enough
(within a few Myrs), to start accreting gas while this gas was still
present in the disk, and to stop the Asteroid Belt from forming a
planet. On the other hand, the spread in ages for some inclusions in
meteorites (CAIs and Chondrules) shows that that the process of
putting together planetesimals must have taken at least 4-6 Myrs
\citep{brearley98:_chond_meteor}.  The pioneering work of
Weidenschilling
(\citeyear{weidenschilling:1977,1980Icar...44..172W,1984Icar...60..553W})
considered the main processes contributing to particle growth,
i.e.~the collisons between grains caused by Brownian motion, vertical
settling, radial drift and turbulence.  Many details in the process
have since then been refined, such as the detailed physics of the dust
sublayer forming at the midplane of the disk \citep{dubmorster:1995},
the trapping of particles in anti-cyclonic vortices
\citep[e.g.][]{klahrhenning:1997,cuzzidobrchamp:1993}, the influence
of aggregate shape on the timescales \citep{1997LPI....28.1517W}, as
well as theoretical \citep{rolling,sliding,coagu} and experimental
\citep[e.g.][]{2000Icar..143..138B} studies of the strength and shape
of the aggregates formed.

One of the interesting results is a time-scale problem related to the
radial drift of particles through the disk.  Once a particle becomes
of the size of cm or larger, it tends to decouple from the gas and
move on Keplerian orbits around the star.  But the gas in the disk
rotates with a slightly sub-Keplerian speed around the star, due to
the small radial pressure support within the disk.  The velocity
difference between the gas and the particle causes friction and
removes angular momentum from the dust particle. This results in a 
systematic drift inward toward the star. The time scale for this
radial drift can be relatively short, on the order of 100 years for
m-sized particles at 1\,AU \citep{weidenschilling:1977}.  Only once
the particle has grown to kilometer size this radial drift stops
because the friction has decreased sufficiently.  The question is
therefore: how can we grow grains from cm to km size within only
$10^3$ orbits?  Normal grain coagulation processes may or may not be
fast enough, but particles of such large size have poor sticking
properties.  So the basic issue of grain coagulation is therefore how
we can speed up coagulation in the models, and how can we enhance the
sticking efficiency for larger particles.

A completely different set of \cdrev{constraints} on grain growth in
protoplanetary disks can be derived from the observations of
circumstellar disks around young stars which are probably in the
process of forming planets right now.  The developments in infrared
astronomy in recent years have opened up a new window to planet
formation: the direct study of protoplanetary disks around T Tauri
stars, Herbig Ae/Be stars and even around Brown Dwarfs. The ISO
satellite, for instance, has provided infrared spectra of Herbig Ae/Be
stars with unprecedented accuracy (e.g.~Malfait et
al.~\citeyear{malfaitwaelk:1999}; van den Ancker et
al.~\citeyear{vdanckerbouw:2000}), giving clues to the composition and
size of dust particles in these disks (Bouwman et
al.~\citeyear{bouwkotanckwat:2000}) and to the geometry of these disks
(Meeus et al.~\citeyear{meeuswatersbouw:2001}).  The new Spitzer Space
Telescope will do the same for T Tauri stars and Brown Dwarfs.  With
infrared interferometry (IOTA, PTI, VLTI, Keck etc) much has been, and
will be learned about the dust mineralogy and structure of the disk at
sub-AU scales (e.g.~Eisner \citeyear{eislanake:2003}) and in the
habitable zones around these stars (van Boekel et
al.~\citeyear{vboekelcryst:2004}).  All of these observations have
drastically increased our knowledge of the physics of the birthplaces
of planets. One of the main discoveries from (sub-)mm observations
(e.g.~Testi et al.~\citeyear{testnat:2003}) and from mid-infrared
spectroscopy (e.g.~van Boekel et al.~\citeyear{vanboekelwaters:2003})
is the ample evidence for grain growth in disks.  However, the
information about grain growth is encoded in a complicated way in the
data we can obtain.  The particles seen by different observations are
different in size, and are located at different locations in the disk.
Extracting the desired constraints on grain growth processes therefore
requires self-consistent modeling of the growth processes with disk
structure and radiative transfer.  For the earliest phases (collapse
and early disk formation), this problem has been addressed by
\citet{1999ApJ...524..857S,2001ApJ...551..461S}.  However, most
observed star+disk systems have already long evolved beyond this very
early phase. Also, the process of the formation of planets may span up
to 10 Myrs or more, which is much longer than the duration of the
early disk phases studied by Suttner et al.

It is the purpose of this paper to confront self-consistent models of
grain coagulation and settling in protoplanetary disks with the
infrared observations of these objects.  Interestingly, this has never
been done before.  In this paper we solve the coagulation equation
(often referred to as the Smoluchowski equation), coupled to the
equation for grain settling and vertical turbulent mixing. The
coagulation equation evolves the dust size distribution in time, while
the settling/mixing equation solves the vertical motion of the dust
particles. At given time intervals we compute the spectral energy
distribution (SED) and images of the disk using a 2-D axisymmetric
continuum radiative transfer code.

We will show in this paper that we are confronted with a new time scale
problem, quite contrary to the one discussed above: we find that the {\em
coagulation of small particles is too fast} to be 
consistent with infrared observations. From
infrared spectroscopy in the 10 micron regime it seems that grain growth
from 0.1 $\mu$m to 2 $\mu$m happens over a time scale of a few $10^6$ years
(van Boekel et al.~\citeyear{vboekelmin:2004}). We will show with our models
that it is very difficult to prevent the complete depletion of grains up to
$100$ $\mu$m within only $10^4$ years. We will suggest that aggregate
fragmentation may provide a piece of the puzzle.  In doing so, we will
necessarily repeat computations done before for the solar system, but
we will do so with better resolution and with considering
observational consequences.

The structure of this paper is as follows: In Sect.
\ref{sec:equations} we describe the equations used to solve the
problem of coagulation in a protoplanetary disk.  From there we build
our model step by step, adding the different processes one at a time,
in order to show the relative importance, and to demonstrate the
solidness of the main conclusion.  In Sect.~\ref{sec-safronov}, we
first briefly discuss the fate of a single particle as it settles to
the disk midplane and grows by sweeping up other grains.  
In Sect.~\ref{sec-models-slice} we solve the coagulation-settling equations for
the entire ensemble of grains in a vertical slice in a disk, producing
time-dependent grain size distribution functions as a function of height
above the midplane.
In Sect.~\ref{sec-globmodels} we use these local
models to build a full disk model and to compute SEDs, with
the conclusion that full coagulation depletes small grains much too
rapidly in order to be consistent with infrared observations of disks.
In Sect.~\ref{sec-destr} we introduce a possible remedy for this puzzle
by showing that aggregate fragmentation might continuously 
replenish the small grains. Finally, in Sect.~\ref{sec-discussion} we discuss
our results.

\section{Equations}
\label{sec:equations}
The distribution of grains of a certain mass $m$, at a certain distance $R$
from the star, a certain height $Z$ above the midplane and at a time $t$ is
given by the distribution function $f(R,Z,m,t)$. It is defined such that
$f(R,Z,m,t)\,m\,dm$ is a dust mass density in g/cm$^3$. The dust grains
undergo several processes. First of all, they settle toward the midplane,
and at the same time they get turbulently stirred back up again. In the
absence of coagulation, an equilibrium solution is eventually reached in
which the upper layers of the disk are devoid of dust of a particular size,
while below a certain $Z$ the grains are fully mixed with the gas and hence
have more or less constant abundance (e.g.~Dubrulle et
al.~\citeyear{dubmorster:1995}; Takeuchi \& Lin
\citeyear{takeuchilin:2002}). This process of settling and stirring has
profound effects on the appearance of protoplanetary disks
\citep{duldomsett:2004}.

As the settling and turbulent stirring process takes place, the grains also
coagulate. The relative motions between the dust grains, necessary for them
to meet each other and form aggregates, are produced by various processes.
In the models presented in this paper we include Brownian motion,
differential settling and relative motions caused by turbulence. These are
the most important processes for coagulation up to cm size particles, and 
we will describe in detail below how they are implemented in our model.

The problem of dust settling and stirring is a problem in the coordinate $Z$
while coagulation is a problem in the coordinate $m$. In principle they have
to be solved simultaneously. In practice, the settling and mixing are solved
in the same subroutine, and the coagulation in another subroutine. By using
the technique of {\em operator splitting} one can switch between the two
subroutines at each time step, thereby solving the simultaneous set of
problems.

\subsection{Cross sections of particles}

An important quantity that will enter the equations is the cross
section of a particle, for collisions with gas particles and with
other particles.  If we assume solid spheres, the collisional cross
section for grain-gas collisions is simply given by $\siggas=\pi a^2$
where $a$ is the radius of the particle.  For collisions between two
grains, the cross section will be $\sigcoag = \pi (a_1+a_2)^2$.
However, in reality, grains formed by aggregation are never compact
spheres.  The will have internal structure which depends on the
formation mechanism.  One extreme of the possibilities are
Particle-Cluster Aggregates (PCA) which are formed when an aggregate
grows by addition of small grains only.  Such particles will tend to
be spherical and porous, with a limiting porosity for large particles
of about 90\%.
The other extreme are Cluster-Cluster-Aggregates (CCA) in
which growth of an aggregate is dominated by accreting aggregates of
its own size.  Obviously, many other possibility exist, including
simultaneous mixture of PCA and CCA (growing by both small and large
aggregates at the same time) and hierarchical growth by switching from
one growth process to another, possibly several times.  The purpose of
this paper is not an in-depth study of these processes, but will be
subject of a future study.  For the current study, we only use the
three classical cases: compact, PCA and CCA particles.  For each of
these growth classes, a unique relation between mass and average
collision cross sections can be derived.  We here use the analytical
fits by \citet{ossenkopf:1993}.  These relations directly provide
$\siggas(m)$ and $\sigcoag(m_1,m_2)$.

\subsection{Settling and vertical mixing}
The settling and mixing of grains is a 1-D time-dependent problem in $z$,
which has to be solved for each $R$ and $m$. We describe the settling and
vertical mixing equations following the discussion of Dullemond \& Dominik
(\citeyear{duldomsett:2004}). We refer to that paper for details. The
equilibrium settling velocity for particles smaller than about 1 cm (Epstein
regime) is:
\begin{equation}
\label{eq-settvelo}
\vset = - \frac{3\Omegak^2z}{4\rhogas\cs}\frac{m}{\siggas} 
\comma
\end{equation}
where $\Omegak\equiv\sqrt{G M_{*}/R^3}$ is the Kepler rotational frequency,
$\cs$ is the isothermal sound speed.
$\siggas$ is the collisional cross-section of the dust grain for
collisions with gas or very small dust particles, i.e. the projected
surface of the grain, averaged over all directions.

The diffusion constant for turbulent mixing $D$ is given by
\begin{equation}
D = \frac{\alphaturb \cs^2}{\Omegak\Sc}
\comma
\end{equation}
where $\alphaturb$ is the turbulence parameter.
The symbol $\Sc$ is the Schmidt number
defined as $\Sc\equiv 1+\St$ with the Stokes number $\St$ given by
\begin{equation}
\St = \frac{3}{4}\frac{m}{\siggas}\frac{\Omegak}{\rho\cs}\alphaturb^{2q-1} 
\fullstop
\end{equation}
In this above equation $q$ is a parameter characterizing the turbulence.
We take it $q=1/2$ in this paper \citep{schreaphenn:2004}.

With the vertical settling velocity and the turbulent mixing constant
we can now define the settling/mixing equation as follows:
\begin{equation}
\left.\frac{\partial f(m)}{\partial t}\right|_{\mathrm{sett}} = 
-\frac{\partial(f\,v_{\mathrm{sett}})}{\partial z} + 
\frac{\partial}{\partial z}\left(D(z)\rho
\frac{\partial (f/\rho)}{\partial z}\right)
\fullstop
\end{equation}
In this paper we solve this equation for each radius $R$ separately. We do
not couple these radii, i.e.~particles are not allowed to move from one
vertical slice to another. In principle this should be included, since
radial drift can be very important. But for simplicity, and since we are
mainly interested in the growth from 0.1 $\mu$m grains to 1 cm grains, we
ignore this process and treat every radius as a separate 1-D vertical
settling/mixing problem, coupled to the coagulation equations described
below.

We solve the settling/mixing equation numerically using implicit
differencing. This is necessary since the Courant-Friedrich-Lewy condition
for the vertical mixing would require much too small time steps for an
explicit solution, and would make the simulation prohibitively slow.

\subsection{Coagulation}
The coagulation of grains is a 1-D time-dependent problem in $m$, which has
to be solved at each $R$ and $Z$. The coagulation equation is
(Schumann \citeyear{schumann:1940}; Todes \citeyear{todes:1949};
Safronov \citeyear{safronov-book}):
\begin{equation}
\label{eq-smoluchovski}
\begin{split}
\left.\frac{\partial f(m)}{\partial t} \right|_{\mathrm{coag}} =   &
\int_{0}^{m/2} f(m')f(m-m')\sigcoag(m',m-m')\\
 &\quad\quad\quad\times\Delta v(m',m-m') dm'\\
&-\int_{0}^{\infty} f(m')f(m)\sigcoag(m',m)\\
 &\quad\quad\quad\times\Delta v(m',m) dm'
\end{split}
\comma
\end{equation}
which is the continuous form of the Smoluchowski equation
(Smoluchowski \citeyear{smoluchowski:1916}). The first term on the
right hand side represents the gain of dust matter in the mass bin $m$
by coagulation of two grains of mass $m'$ and $m-m'$. The second term
represents the loss of dust matter in the mass bin $m$ by coagulation
of a particle of mass $m$ with a particle of mass $m'$.  $\Delta
v(m_1,m_2)$ denotes the average relative velocity between these two
particles. This average velocity may consist of random motions but
also of systematic drift between particles of different mass. The
combination $K(m_1,m_2)\equiv \sigcoag(m_1,m_2)\Delta v(m_1,m_2)$ is
called the {\em kernel} of the coagulation equation.

We include three processes leading to a $\Delta v(m_1,m_2)$: Brownian motion,
differential settling and turbulence. For Brownian motion, the average relative
velocity is given by:
\begin{equation}
\Delta v_{\mathrm{b}}(m_1,m_2) = \sqrt{\frac{8kT (m_1+m_2)}{\pi m_1 m_2}}
\fullstop
\end{equation}
This represents an average of random velocities, which is highest when
both particles have the smallest mass. Therefore, Brownian motion
favors collisions in which at least one collision partner has low
mass.  Once these lowest mass particles get depleted, the next higher
mass particles (which are aggregates of the smallest particles) will
coagulate etc. Brownian motion will therefore lead to a narrow size
distribution which slowly moves to larger sizes. This
hierarchical growth procedure leads to aggregates with a fractal
structure: so-called cluster-cluster aggregates (CCA) \citep{meakin87}.

Differential settling is the process by which large grains, which settle
faster than small grains, sweep up the smaller grains on their way to the
midplane. 
This process is very similar to the formation of rain drops in clouds
in the earth atmosphere (`rain out'). The systematic relative velocity is:
\begin{equation}
\Delta v_{\mathrm{s}}(m_1,m_2) = |\vset(m_1)-\vset(m_2)| \quad .
\end{equation}
This is clearly zero for particles with equal $\siggas/m$ ratio.  In
the special case of a uniquely defined $\siggas/m$ for any given $m$, 
as we are using
in this paper, this means that the relative velocity is zero for equal
mass particles, since they both settle at equal speed. The
differential settling as a source of collisions works best for
particles of very different $\siggas/m$ (and therefore mass).
Therefore typically the largest mass aggregates will sweep up the
smallest particles.  This leads to relatively compact particles with
porosities typically of the order of 90\%.  Such
aggregates are often called particle-cluster aggregates (PCA).

Turbulence-driven coagulation is a rather complex process. It requires
detailed calculations of the statistics of motions of particles (Voelk
et al.~\citeyear{voelkmorroejon:1980}).  Unfortunately, these
calculations are too complex to directly build into a coagulation code
like the one we present here.  \citet{1984Icar...60..553W} has derived
fitting formulae to the results of Voelk et al., which we implement
with slight modifications here.  We refer a discussion of these to
appendix \ref{app-voelk}.  It should be noted that these formulae
introduce a considerable uncertainty into the present coagulation
code, since they may depend on details of the turbulence which are not
well known.  For example, turbulence driven by the magneto-rotational
instability \citep{balbushawley:1991} may have different properties
than the turbulence assumed by Voelk et al..  \cdrev{The results we
  focus on in the current paper, which is the fast disappearance of
  small grains, is present even if we ignore turbulence as a source of
  relative velocities.  Therefore, the uncertainties introduced by
  assuming a particular turbulence spectrum are acceptable.}

Solving Eq.~(\ref{eq-smoluchovski}) numerically is challenging.  In
appendix \ref{app-numerics} we describe our numerical algorithm, and
in appendix \ref{app-tests} we show the results of a comparison
against a test case described in the literature.

\subsection{Grain fragmentation}\label{subsec-destr}
In most of this paper we shall ignore aggregate fragmentation/destruction, 
since we
are interested to see what happens to the observables of the
protoplanetary disk when coagulation is able to develop to its full
extent. But in Section \ref{sec-destr} we will present a
single-vertical-slice calculation of a situation in which grain
fragmentation is included. Including grain fragmentation in a proper way
is challenging. But since the purpose of that section is only to
demonstrate the main effects of replenishment of the smaller grains,
we are satisfied with an {\em extremely} simplified implementation of
fragmentation. We assume two colliding aggregates to disintegrate
entirely into monomers if their collision energy, when divided over
the sum of the masses of the two particles, exceeds a certain value.
As our fragmentation energy we choose 1250 erg/g, which, for equal mass
particles, amounts to a collision velocity of 1 m/s. This represents
rather loosely bound particles, which enhances the effect of
fragmentation. In reality somewhat higher fragmentation velocities may be
required, and also other processes such as partial fragmentation (or
cratering). But for the purpose of demonstration of principle this
very simple recipe is justified.

The recipe described above is implemented in the following way. The mass
bins below $a\lesssim 0.5 \mu$m are considered to be `monomer bins'. Any
collision between particles of size larger than $0.5\mu$m with a collision
energy (divided by mass) larger than the critical value will remove the mass
from the two original bins and return this mass into the monomer bins. The
distribution in which it is returned is fixed, and represents the starting
distribution shape (MRN, see below).

\section{Prelude: the one-particle model}\label{sec-safronov}
Before we present our full-fledged coagulation-settling-mixing models
we briefly revisit the simple one-grain model discussed by Safronov in
his book (\citeyear{safronov-book}), since this model is very
illustrative of the phenomena presented in the subsequent sections.
The one-particle model follows the growth and settling of a single
dust particle in a disk, assuming that all other dust particles remain
suspended in the disk and do not coagulate. As the particle settles,
it sweeps up the small grains suspended in the disk. Therefore, the
mass of the particle is an increasing function of time: $m(t)$ and the
height of the particle above the midplane $z(t)$ decreases. Here, and
in the remainder of the paper, we take a very simple disk vertical
structure in order not to complicate matters more than necessary. We
assume that the disk is vertically isothermal and the gas density
obeys hydrostatic equilibrium.  We therefore have:
\begin{equation}
\rhogas(z) = \frac{\Sigma}{\sqrt{2\pi}\Hp}
 \exp\left(\frac{-z^2}{2\Hp^2}\right)
\comma
\end{equation}
where $\Hp$ is defined as
\begin{equation}
\Hp= \frac{\cs}{\Omegak} = \sqrt{\frac{kT}{\mu m_p}\frac{R^3}{GM_{*}}} 
\comma
\end{equation}
with $M_{*}$ the mass of the central star, $\mu$ the mean molecular weight,
taken to be $\mu=2.3$ (for molecular gas) and $R$ the distance from the
star.  The temperature $T$ is assumed to be the midplane temperature of a
disk irradiated under an angle of $\alphairr=0.05$ around a star of
$M_{*}=0.5\,M_{\odot}$, $T_{*}=4000$ K and $R_{*}=2.5\,R_{\odot}$.
\keesrev{If we assume an isothermal disk structure without a warm surface
layer, then the temperature of such a disk is $T= \alphairr^{1/4}
\sqrt{R_*/R}\,T_{*} =204\,$K}.

Now if we let the particle in question settle toward the midplane according
to the settling velocity Eq.~(\ref{eq-settvelo}), then the equation of
motion for that particle becomes:
\begin{equation}
\frac{dz}{dt} = \vset = -\frac{3\Omegak^2z}{4\rhogas\cs}\frac{m}{\siggas} 
\fullstop
\end{equation}
At the same time, the mass of the particle increases according to
\begin{equation}
\frac{dm}{dt} = 0.01\, \rhogas(z) |\vset| \sigcoag(m)
\comma
\end{equation}
where the factor $0.01$ is the dust-to-gas ratio. The above two equations
form a coupled set of ordinary differential equations which can be
integrated using standard integration schemes.  Since the evolving
particle quickly is much larger than the particles suspended
in the gas, we assume $\siggas=\sigcoag$ - but we will see below and
in section \ref{sec-models-slice} that this is not fully correct.

\begin{figure}
\centerline{\includegraphics[width=9cm]{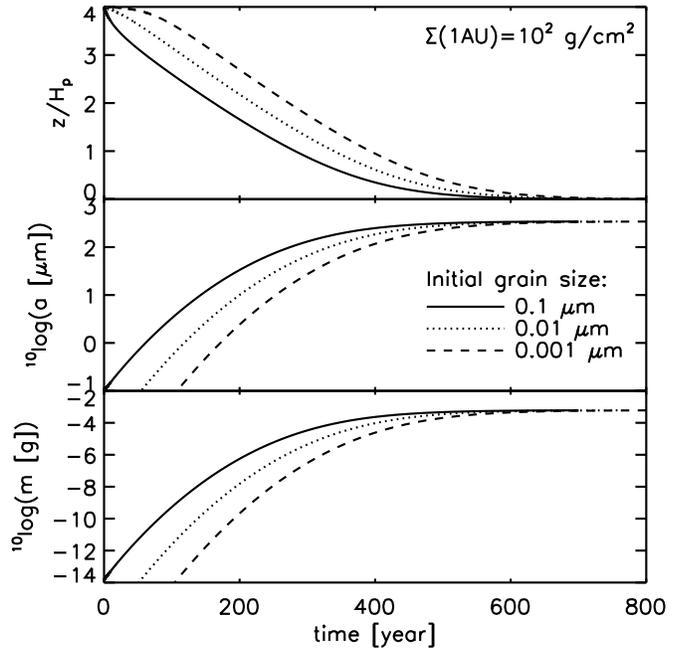}}
\caption{\label{fig-oneparticle-a}The time-evolution of the height $z$, radius
$a$ and mass $m$ of a dust grain in the simple one-particle model, for 
three different initial dust radii. The specific weight of the dust is
$3.6$ g/cm$^3$.}
\end{figure}

In Fig.~\ref{fig-oneparticle-a} we show the resulting $z(t)$, $m(t)$ and
$a(t)$ assuming spherical compact silicate grains in a disk with
$\Sigma_{\mathrm{gas}}=10^2$ g/cm$^2$ \keesrev{at $R=1\AU$},
starting at a height $z=4\,H_p$ (with $H_p$ the pressure scale height of the
disk given by $H_p=\sqrt{kTR^3/\mu m_p G M_{*}}$), and with grains of
different initial radius (mass).
As one can see from these figures, the grain grows exponentially as
it sweeps up matter during its decent. It reaches the midplane as a cm size
pebble in a few hundred years, even though it would have taken a few million
years to reach the midplane if the grain would not have grown to larger size
on its way to the midplane. Interestingly the time it takes to reach the
midplane is almost independent of the initial grain size. 
\begin{figure}
\centerline{\includegraphics[width=9cm]{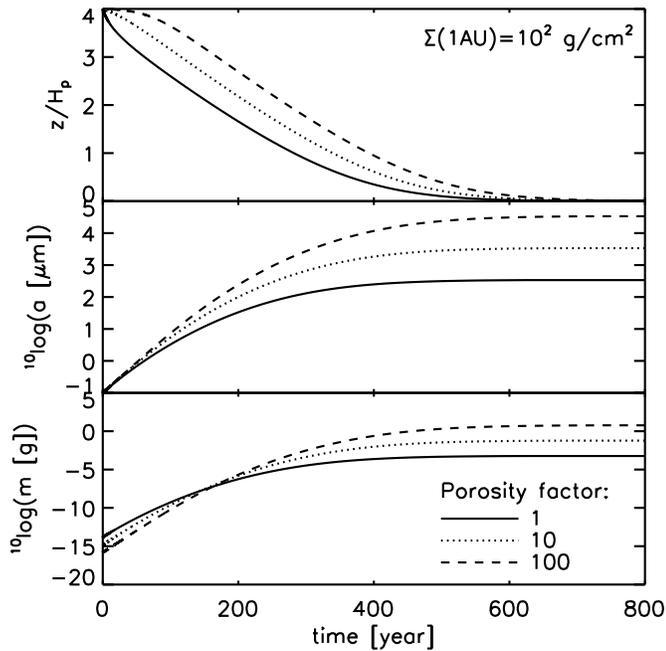}}
\caption{\label{fig-oneparticle-p}As Fig.~\ref{fig-oneparticle-a}, but
now for different porosities of the grains.}
\end{figure}
In fact, it is also almost independent on the compactness of the
grains, as can be seen in Fig.~\ref{fig-oneparticle-p}. A porous grain
has lower settling velocity but larger cross-section. It therefore
sweeps up more matter, allowing it to regain the velocity it would
have had when it would have been more compact
\citep{1997LPI....28.1517W}.  Moreover, the porous grain ends up at the
midplane with a larger mass than a compact grain. If one were to redo
the experiment with fractal grain growth (resulting in cluster-cluster
aggregates) the resulting end size of the grain diverges completely,
becoming (theoretically) larger than the entire disk but with
virtually infinite porosity. This is clearly unphysical.  In reality,
at some particle size a transition to more compact particles with
fractal dimension 3 will take place, avoiding this divergence.  At
what size and how this happens is still one of the main questions in
the study of planet formation.

As a preliminary conclusion from this simple exercise one can say that
grain growth due to this `raining effect' happens on a very short time
scale, much shorter than the typical life time of protoplanetary disks. And
it seems that porousness or fluffiness of the grain does not slow down this
growth process. It merely increases the final grain mass.

\section{Local disk models with coagulation, settling and mixing}
\label{sec-models-slice}
The result of the previous section has shown that coagulation can proceed
very quickly through the differential settling process. However, that
calculation was based on the simple assumption that a single grain follows
this mode of growth while all other grains remain suspended in the gas. In
reality this is not the case. All grains evolve simultaneously, and settle
simultaneously. It is not clear at which height above the midplane this
`rain-out' process starts, and what the effect of the collective growth
is. Moreover, as will be shown below, the initial mode of growth is Brownian
motion, not differential settling, and only after a certain time the
differential settling will take over.

In this section we will show the results of the full-fledged
coagulation-settling-mixing calculations following the equations outlined
above. We do the calculation first for a single vertical slice \keesrev{at
$R=1\AU$}. As our disk vertical structure we adopt the same structure as in
Section \ref{sec-safronov}. Our initial distribution function is a
\citet{mrn:1977} (MRN) distribution from 0.1 $\mu$m to 0.5 $\mu$m.

In order to show all the effects more clearly, we proceed in steps. First we
assume no settling, nor mixing nor coagulation by differential settling nor
coagulation by turbulence: we only include Brownian motion (model S1).  Then
we show a model in which we include the raining effect as well (the
coagulation by differential settling), but still no vertical mixing (model
S2). We then also include turbulent mixing (model S3). Finally we also
include coagulation by turbulence (model S4). These models will demonstrate
the speed at which the coagulation takes place.

\begin{table}
\centerline{
\begin{tabular}{c|ccccc}
   & Brownian & DiffSett & TurbMix & TurbCoag & Poros \\
\hline
S1  & $\surd$ &         &         &         & Comp \\
S2  & $\surd$ & $\surd$ &         &         & Comp \\
S3  & $\surd$ & $\surd$ & $\surd$ &         & Comp \\
S4  & $\surd$ & $\surd$ & $\surd$ & $\surd$ & Comp \\
S5  & $\surd$ & $\surd$ &         &         & PCA \\
S6  & $\surd$ & $\surd$ &         &         & CCA 
\end{tabular}}
\caption{\label{tab-models-s}Table of parameters of single vertical slice
models (the S-series). First column: model name, second column: Brownian
motion, third column: differential settling (`rain effect'), fourth column:
turbulent mixing, fifth column: turbulence-driven coagulation and sixth
column: porosity (Compact, Particle-Cluster-Aggregate or
Cluster-Cluster-Aggregate).}
\end{table}

The models S1, S2, S3 and S4 are for compact spherical silicate grains. In
order to show what the effects of non-compactness would be, we also present
models similar to S4, but with particle-cluster aggregates (PCA) and
cluster-cluster aggregates (CCA). These are the models S5 and S6
respectively. 

\begin{figure*}
\centerline{\includegraphics[width=16cm]{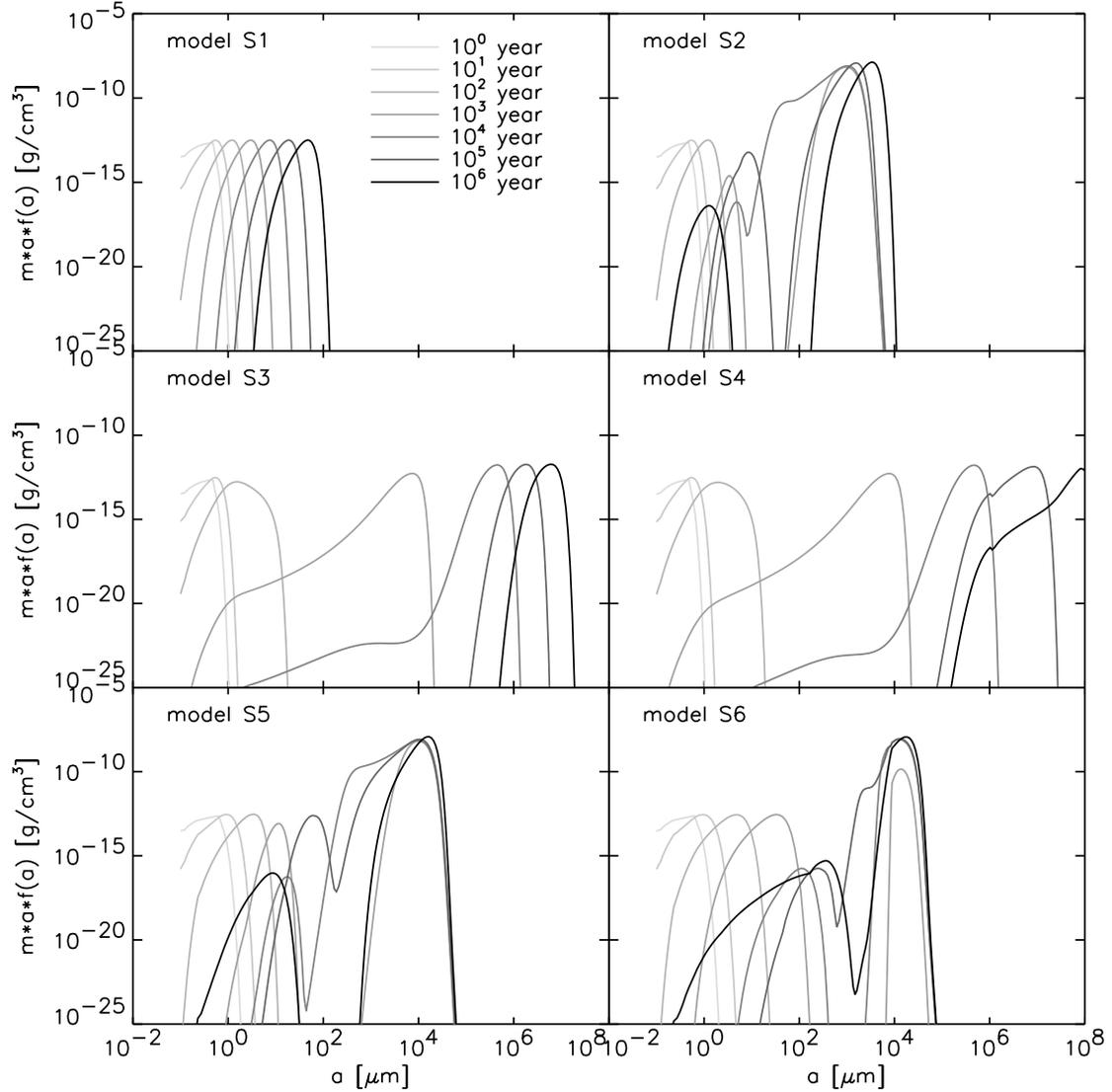}}
\caption{\label{fig-series-s}The time-evolution of the distribution function 
for models S1$\cdots$S6 (see table \ref{tab-models-s} for the model
definitions).}
\end{figure*}

Table \ref{tab-models-s} provides an overview of the models of this
section. The resulting midplane dust distributions at different times
are are shown in Fig.~\ref{fig-series-s}.  In these plots, we show the
actual mass distributions (i.e. $m^2 f(m)$ when plotted over $\log m$
or $m\cdot a\cdot f(a)$ when plotted over $\log a$, like $\nu F_\nu$
for spectral energy distributions).  On the x-axis we deliberately
plot the radius of the grain $a$ instead of the mass $m$, because the
radius of grains is a more familiar quantity. For the PCA and CCA
models (models S5 and S6) this radius is the equivalent radius as if
the grain would have been compact (i.e.~an equivalent radius $a$ for
the PCA/CCA models corresponds to the same particle mass as for the
compact models). For the PCA/CCA models the real radius is much
larger.

\begin{figure*}
\centerline{\includegraphics[width=16cm]{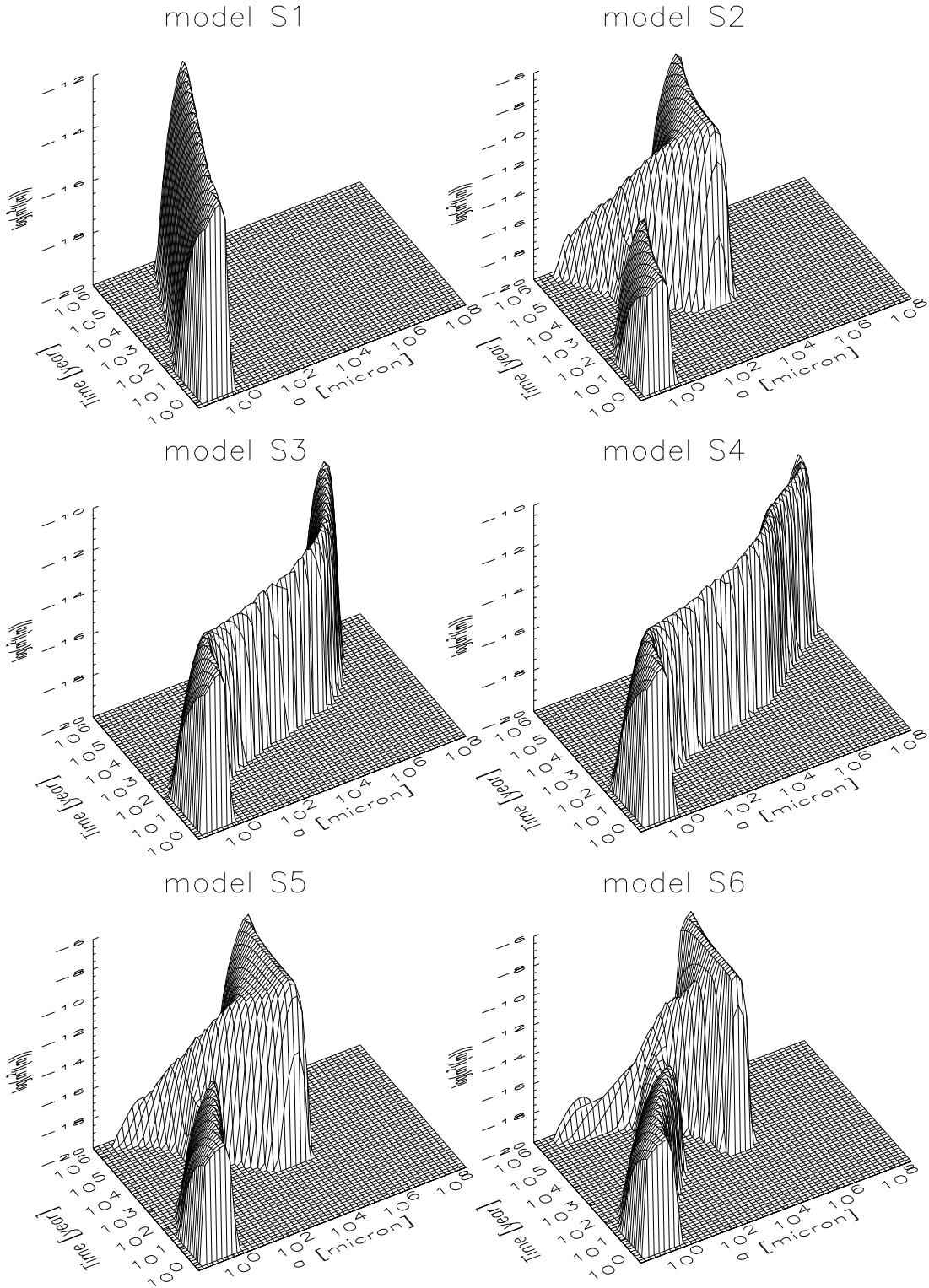}}
\caption{\label{fig-series-s-3d}Like figure \ref{fig-series-s}, but in a
  2d view.}
\end{figure*}

For clarity, we plot the results in two different ways.  Figure
\ref{fig-series-s} shows the midplane size distributions at different times for
the various models in a normal 2D plot.  Figure~\ref{fig-series-s-3d}
shows the same results in a 3D way, i.e. we provide a separate axis
for time.  This is useful in particular for the calculations in which
the size distribution develops two different peaks - such development
gets confusing in the 2D plot.  We plot $m\cdot a\cdot f(a)$ as a function of
$\log a$ so that surface under the distribution function corresponds to total
density of dust at that position in the disk.

As one can see from the upper left panels of Figs.~\ref{fig-series-s}
and \ref{fig-series-s-3d}, the coagulation by Brownian motion tends to
produce a peak distribution with a certain width.  The peak moves
toward larger sizes in a self-similar way.  But the speed with which
it moves toward larger sizes is proportional to
$a_{\mathrm{peak}}\propto t^{2/5}$. This can be understood in very
simple terms.  If we assume for simplicity that we have to deal with
a mono-disperse size distribution of compact particles with size
$m(t)\propto t^{\mu}$, it is clear that the number density of
particles is proportional to $n(t)\propto m(t)^{-1}\propto t^{-\mu}$.
The relative velocity is $v(t)\propto m(t)^{-1/2}\propto t^{-\mu/2}$.
For the compact particles in this calculation, the collisional cross
section is $\sigma(t)\propto m(t)^{2/3}\propto t^{2\mu/3}$.  The
change in mass of the particles is given by $\mu
t^{\mu-1}=\frac{dm}{dt}\propto m(t)n(t)\sigma(t)v(t)$.  Comparing the
powers to which $t$ is raised in this equation we easily find
$\mu-1=\mu-\mu+2\mu/3-\mu/2$ with the solution $\mu=6/5$ implying
$m(t)\propto t^{6/5}$ and $a(t)\propto t^{2/5}$.  While for the very
smallest grains this is an important growth mechanism, clearly this is
not efficient to grow to very large sizes.

If the differential settling is included (model S2) the initial stage
of growth at the midplane is still Brownian motion. But at some point
($t\simeq 500$ year) one can see a sudden `rain shower' appearing
around grain sizes of $a\simeq 1$ mm.  This differential settling has
already started at higher elevations before that time, but has not yet
reached the midplane until $t\simeq 500$ year. 
The abrupt appearance of $a\simeq 1$ mm size
grains at $t\simeq 500$ year means that the `rain drops' have finally
reached the midplane and populate the midplane mass bins of size $a\simeq 1$
mm, giving rise to the isolated peak distribution around this size
seen in the upper-right panel of Fig.~\ref{fig-series-s}. During the
`rain shower' the height of the original peak of the distribution 
function (at smaller grain sizes) is strongly reduced: these small grains
are swept up by the descending larger particles. This process
is similar to what happens in the earth atmosphere when aerosols (smog
particles) are washed out of the sky by rain.

The time
scale for this `rain shower' to reach the midplane is very similar to
that predicted by the one-particle model. The particles are, however,
about 3 times larger than in the one-particle model. This is because
the collective raining allows grains to coagulate with particles of
similar (though not equal) size, thereby increasing the collision
cross-section of the grains. The maximum increase factor would be a
factor of $\sigcoag/\siggas=4$, if the grains were allowed to
coagulate with equal size partners.  But since the differential
settling only works for particles of unequal size, this factor is
reduced somewhat, i.e.~becoming a factor of 3 in our simulation.

It is interesting to see that the peak value of $m\cdot a\cdot f(a)$ of the
rained-down grain population (after 500 years) is much higher than the peak
value of the initial distribution, even though the peak width is not much
different from the width of the original distribution. This may appear as a
violation of mass conservation. The explanation is that due to the settling
virtually all the dust grains larger than about 10 $\mu$m have settled in an
extremely thin midplane layer with very high dust density.  This is the
``dust subdisk'' in which, as is sometimes believed, gravitational
instabilities may trigger planetesimal formation (Goldreich \& Ward
\citeyear{goldward:1973}; Youdin \& Shu \citeyear{youdinshu:2002}). This
thin midplane layer is allowed to form in model S2 because we have
explicitly switched off turbulence.  In practice there is doubt whether
such a thin layer can exist, as even the slightest bit of turbulence 
thickens the dust subdisk \citep{1995LPI....26.1477W}

After the `rain-out' nothing much happens, since the `rain shower' has
cleared out a large fraction of the small grains in the disk, and there is
not enough of them left to continue the raining effect.  Basically, the rain
shower is over. The remaining population of small grains continues to grow
via Brownian motion until it merges in the population of rained-down
grains. It should be noted, however, that a minute amount of small grains
from higher elevations, which have survived the rain shower, still slowly
but surely settles toward the midplane. \keesrev{
These grains reach the midplane during the later stages of the
simulations.  The slightly larger grains arrive first, the smallest
grains last.  Once arrived at the midplane, the grains get
incorporated into larger grains due to Brownian motion. 
The combined effect is a dip in the size
distribution around about a size of 10$\mu$m.  This dip is widening with
time, producing a size distribution with two peaks.  Due to the
arrival of ever smaller grains, the small particle peak moves to the
left, while the large particle peak slowly moves to the right (because of
further Brownian motion grain growth). If the small grains would
not be incorporated into the big grains (for instance, if one would
switch off Brownian motion coagulation), the settled size
distribution would freeze out, with the dip between the two peakes
filled in.
}

Around 1 Myr the secondary peak is seen at 1 $\mu$m. It
appears as if this peak is only 10000 times smaller than the value of
the initial distribution. While this is a strong depletion of small
grains, it may not be enough to render the disk optically thin.
However, here again this is somewhat deceptive due to the thinness of
the dust subdisk. For grains of this size the dust subdisk is 1\% of
the pressure scale height of the disk at 1 Myr. This brings the
depletion factor to about $10^6$.

If we include the vertical turbulent mixing with $\alphaturb=0.01$ (model
S3), then a new interesting phenomenon occurs.  The initial evolution
remains the same as for S2 (Brownian motion followed by raining), but the
differential settling process does not stop anymore. Because the rained-down
grains are now stirred up from the midplane again, they can rain down a
second and third time, continuing to sweep up any grains they meet. This
prolongs the fast growth process considerably, and strongly depletes the
small grains from the disk. This process is \keesrev{somewhat} similar to
the growth of giant hail stones in cumulonimbus clouds in the earth
atmosphere: before such stones reach the earth surface, they often get
transported back to high altitudes by the strong updraft within the
cloud. In this way these hail stones get multiple chances to collect rain
drops and smaller hail stones on their surface, allowing them to grow to
sizes as large as decimeters. \keesrev{The main difference is that the air
movement in cumulonimbus clouds constitutes a systematic flow,
while in our models the turbulent motions are random.}

In model S4 we also include the coagulation by the turbulence itself. It is
interesting to see that this process does not appear to make much difference
in the earlier phases of the coagulation process. The differential setting
effect (prolonged by the vertical stirring) is clearly the dominant process
up to $10^4$ years. After that, the turbulence-driven coagulation takes over
and manages to grow the grains even further. However, by this time we are
well in the regime of boulders, in which many of the equations we use are no
longer valid. Note a tiny wiggle in the distribution function at $10^6$ 
$\mu$m: this is an effect that can be traced back to the discrete jump in
the slope of the fitting formulae of appendix \ref{app-voelk}.

\cdrev{From the above calculations it is clear that, if we allow coagulation
  to work with perfect efficiency (i.e. maximum sticking, no destruction)
  coagulation happens on an extremely short time scale, even if we only
  include coagulation by Brownian motion and differential settling into our
  calculations.  If we would include other processes such as coagulation by
  radial drift, while still assuming perfect efficiency, then this is only
  aggravated instead of alleviated. It seems therefore that this result is
  rather robust.  We shall see in section \ref{sec-destr} that destructive
  collisions may indeed play an important role.}

\begin{figure}
\centerline{\includegraphics[width=9cm]{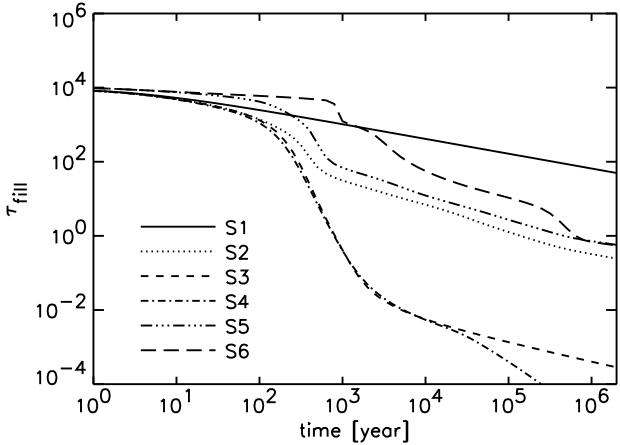}}
\caption{\label{fig-tau}\cdrev{The vertical optical depth in the UV at 1 AU
  (i.e. the integrated projected surface of the dust particles) as a
  function of time for the vertical slice models S1\ldots S6.}}
\end{figure}

\subsection{The optical depth of the models}
A first hint on the effects of the coagulation on the appearance of
a disk can be taken from the optical depth at short wavelength,
i.e. at wavelength corresponding to the stellar photons which are
heating the disk surface and are responsible for the flaring in the
disk.

In Fig.~\ref{fig-tau} we \cdrev{show the vertical} optical depth at UV
wavelengths through the disk.  For the cross section we have taken
the projected geometrical cross section, i.e. $\siggas$.  We can see
that for all models, the optical depth decreases significantly due to
the coagulation.  While in the pure Brownian Motion case S1, the
optical depth decreases by a factor of 100 in 10$^6$yrs, including
settling already decreases the optical depth by more than four orders
of magnitude in the same time, pushing the optical depth below 1.
With turbulent mixing and turbulent coagulation, the effect becomes
enormous, and the disk becomes fully transparent in only 1000yrs.
Using PCA and CCA grains does slow down the process somewhat.  In
particular, in the CCA calculation the optical depth stays constant
for about 1000 years, but then also starts to decrease quickly.
After 10$^6$ years, also in this case the optical depth is below one.
\cdrev{The possibility of change in optical depth on short time scales
  was also indicated by \citet{1980Icar...44..172W,weid1997} who found
  that the Rosseland optical depth may decrease by an order of
  magnitude in about 1000 orbital periods.}

\section{Global disk models with coagulation, settling and mixing}
\label{sec-globmodels}
The short coagulation time scales will have consequences for the infrared
and optical appearance of protoplanetary disks. In this section we glue a
series of vertical slices (annuli) together to form a full disk model.  We
perform the coagulation-settling-mixing calculations in each of these slices
and make snapshots of the distribution function at given times. From this
series of slices we construct a 3-D axisymmetric (i.e.~effectively 2-D) disk
model at each of the snapshot times, and feed these disk models into a
multi-dimensional continuum radiative transfer code called {\tt RADMC} (see
Dullemond \& Dominik \citeyear{duldomdisk:2004}). With this code we can
produce synthetic spectral energy distributions (SEDs) for the disk at the
given times, as well as images at various wavelengths and inclinations.

\begin{table}
\centerline{
\begin{tabular}{c|ccccc}
    & Brownian & DiffSett & TurbMix & TurbCoag & Poros \\
\hline
F1  & $\surd$ & $\surd$ &         &         & Comp \\
F2  & $\surd$ & $\surd$ & $\surd$ &         & Comp \\
\end{tabular}}
\caption{\label{tab-models-f}Same as table \ref{tab-models-s}, but
now for the full disk models F1 and F2.}
\end{table}

Our model has an inner radius $R_{\mathrm{in}}=0.7\,\AU$, outer radius
$R_{\mathrm{out}}=100\,\AU$ and has a gas surface density profile
$\Sigma_{\mathrm{gas}}(R)=\Sigma_0 (R/\AU)^{p}$ with $p=-1.5$ and
$\Sigma_0=400\,\gram/cm^2$, which amounts to a disk with
$M_{\mathrm{disk}}=0.005\,M_{\odot}$. \keesrev{In total we glue 40 vertical
slices together to form the full disk model. The radii of these slices are
log-spaced between $R_{\mathrm{in}}$ and $R_{\mathrm{out}}$, meaning that
the width of each slice is about 14\% of its radius.} We take the
same stellar parameters as in previous sections, and similarly we assume
compact silicate grains.  \keesrev{The temperature of the disk is a function
of radius $R$ only, i.e.~it is constant in vertical direction. As in the
previous sections, the temperature is assumed to follow from thermal balance
assuming that the disk is passive and irradiated by the star under an angle
$\alphairr=0.05$, leading to $T(R)= \alphairr^{1/4} \sqrt{R_*/R}\,T_{*}$.}
We present two models. Model F1 has no turbulent stirring (we set
$\alphaturb=0$) while model F2 has turbulent stirring with $\alphaturb=0.01$
(see table \ref{tab-models-f}).

\begin{figure*}
\includegraphics[width=9cm]{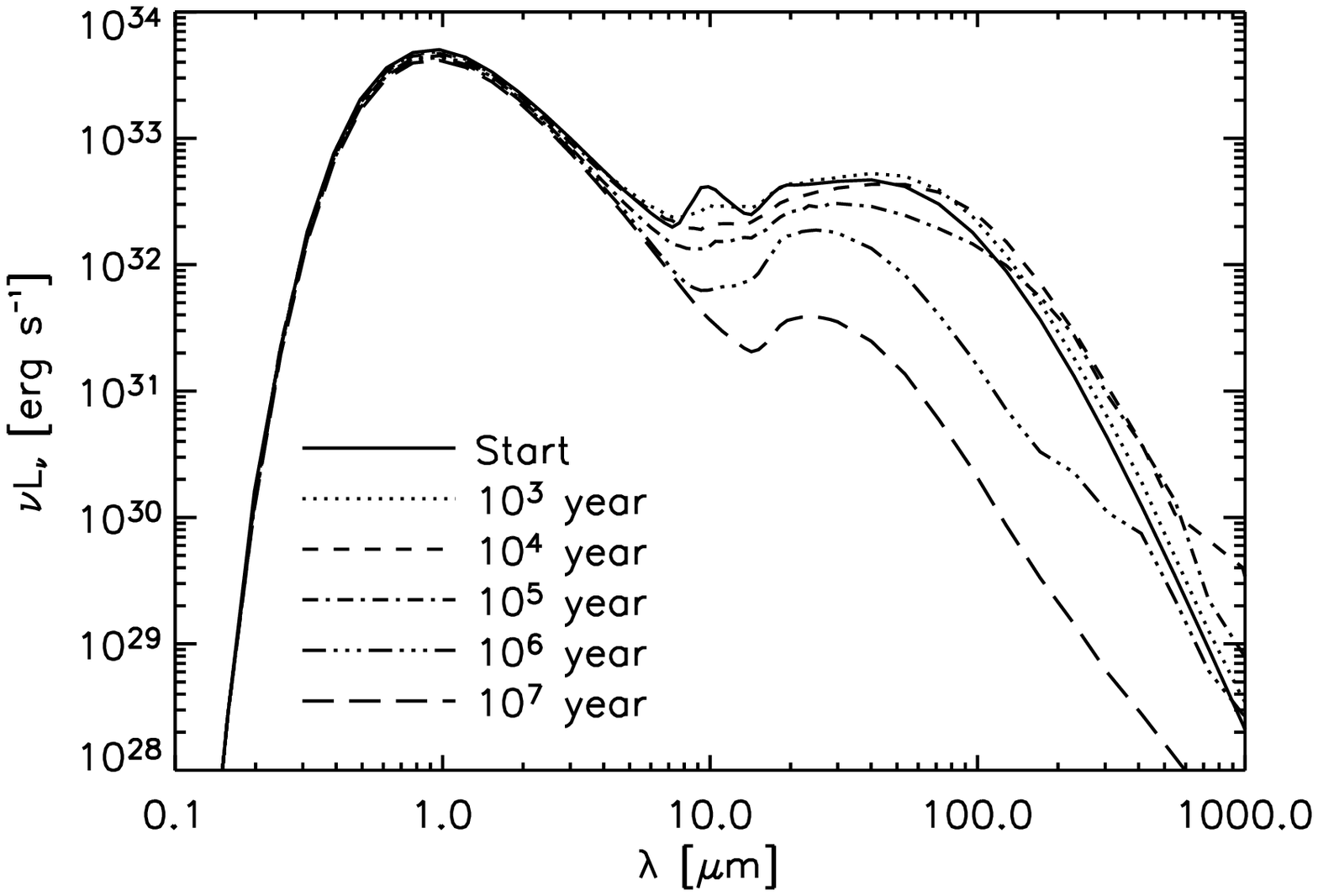}
\includegraphics[width=9cm]{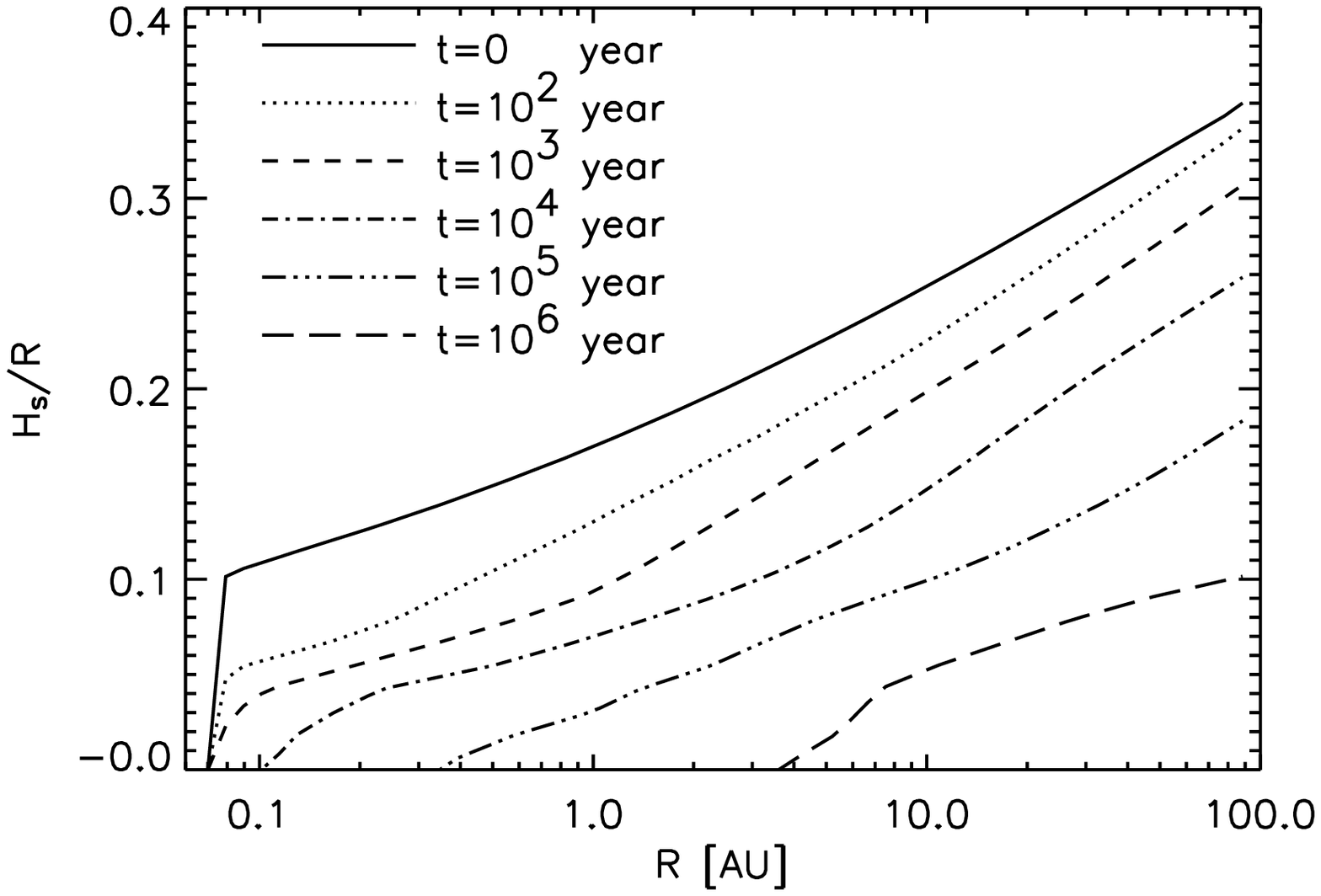}\\
\includegraphics[width=9cm]{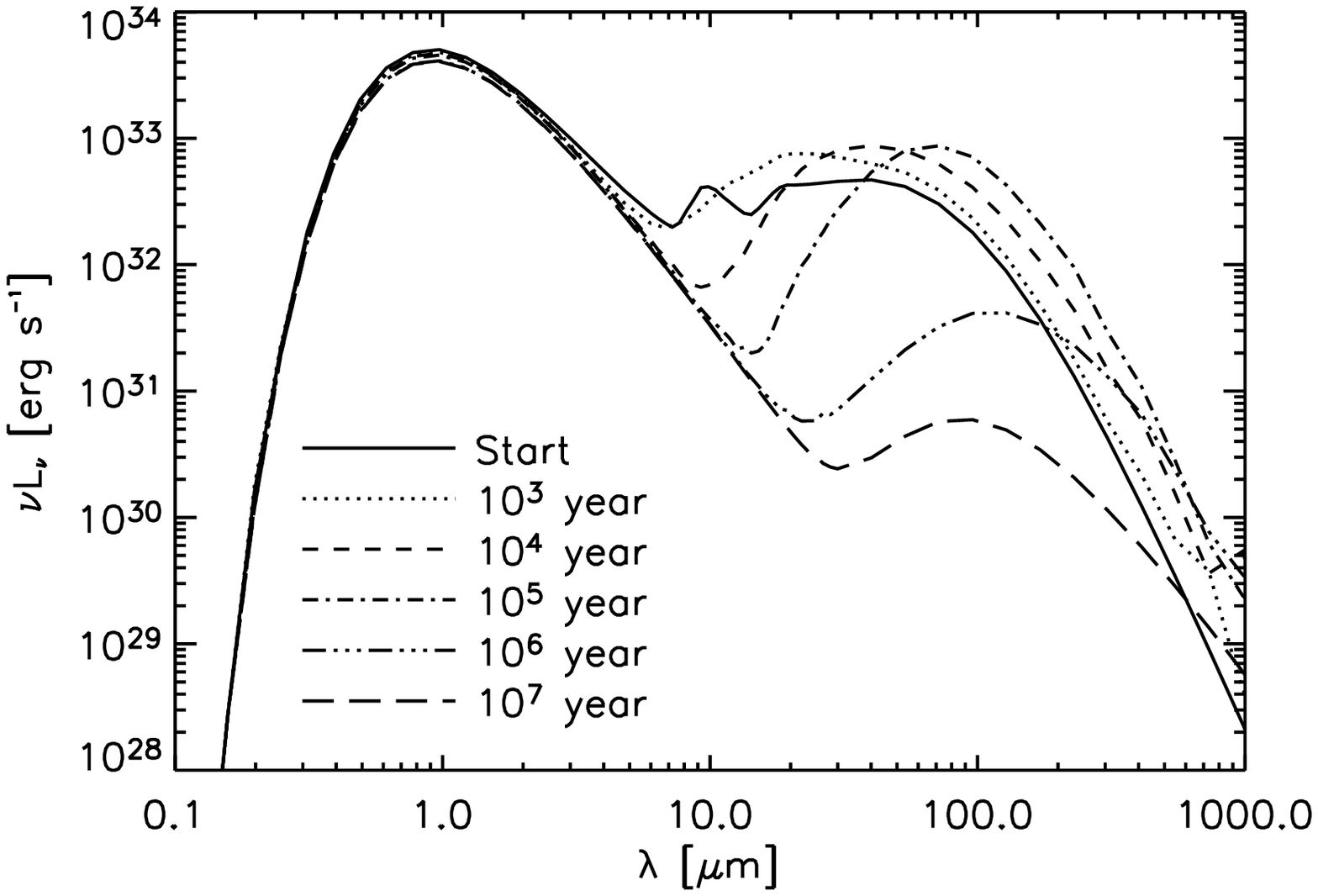}
\includegraphics[width=9cm]{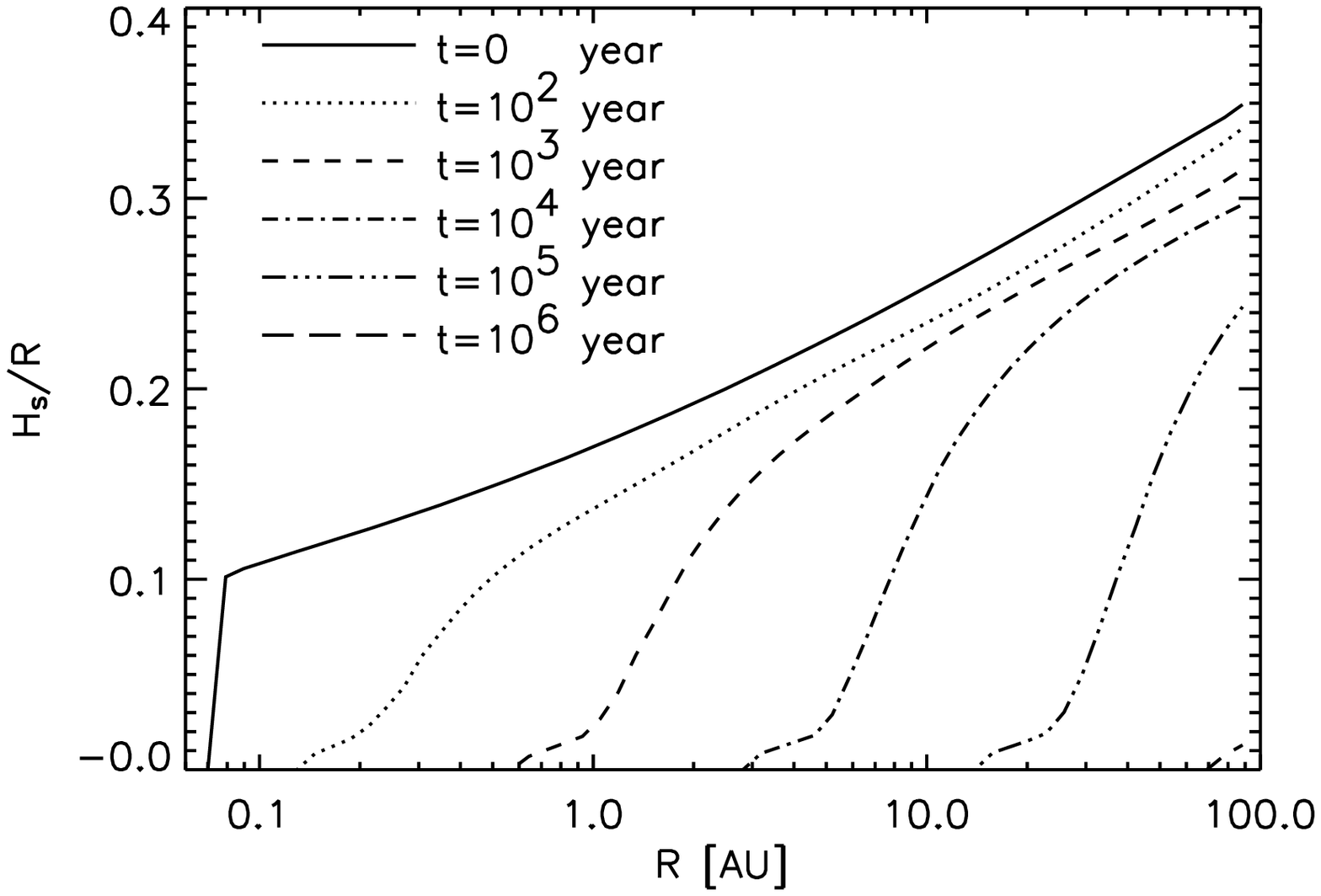}
\caption{\label{fig:fullmodels}\cdrev{The SEDs (left column and surface
  heights (right column) of the full disk models. The two top panels
  show the result for model F1, the calculation without turbulence.
  The bottom panel results are for model F2, including turbulence.
  Different lines indicate the results after different times, as shown
  in the legends.}}
\end{figure*}

The left column of Fig.~\ref{fig:fullmodels} shows the evolution of
the SED of the full disk model without turbulence (F1). One sees that
very early the mid-IR drops, while the far-IR remains. This is because
the coagulation processes are faster at small radii, and therefore the
depletion of small grains happens faster at small radii than in the
outer regions of the disk. It is also clear that the 10 $\mu$m feature
quickly loses strength but does not disappear before most of the
mid-IR continuum flux disappears as well. Its shape also does not
change appreciably. This is an interesting phenomenon, since
observations of the 10 $\mu$m feature of T Tauri stars and Herbig
Ae/Be stars often show flattening and weakening of this feature which
is generally interpreted as a signature of grain growth (van Boekel et
al.~\citeyear{vanboekelwaters:2003}; Przygodda et
al.~\citeyear{przygoddaboek:2003}; Meeus et
al.~\citeyear{meeussterz:2003}; Honda et
al.~\citeyear{hondakataza:2003}).  According to the present
calculations such flattened features are not predicted.  Moreover, the
entire mid-IR flux vanishes much too quickly (well within $10^5$
years). In fact, by $10^6$ years most of the IR excess (also far-IR)
has vanished, which is clearly inconsistent with observations of T
Tauri stars and Herbig Ae/Be stars.
\cdrev{The same behavior is reflected by the surface height in the disk,
plotted in the upper right panel of figure \ref{fig:fullmodels}.  The
surface height is calculated by following starlight radially away from
the stellar surface and determining at what location an optical depth of unity
for a wavelength of 0.55$\,\mu$m is reached.  In contrast to the vertical
optical depth shown for the slab calculations (see
figure~\ref{fig-tau}), this plot is also sensitive to the vertical
distribution of the material.  It measures both loss in total optical
depth, and settling of the opacity carriers toward the midplane and
therefore is the most important indicator for understanding the SED.
The surface height initially begins to decrease globally, which is
mainly due to the settling motion of particles in the uppermost
layers \citep{duldomsett:2004}.  In the inner disk regions, the
effects of coagulation quickly lead to a transparent disk, with the
disk ``surface'' disappearing.  The boundary where photons are
intercepted moves outward and reaches 30 AU in 10$^6$ years, while in
the outer disk, the surface height continuously decreases.
If one includes turbulence in the model, then the SEDs become as
shown in the two bottom panels of Fig.\ref{fig:fullmodels}. In this case the
problem is even more acute.}

It is interesting to see that as the mid-IR flux vanishes, the far IR flux
temporarily increases. This is because of energy conservation. The dust
coagulation happens much faster at small radii than at large radii.  As the
inner regions of the disk become optically thin, the outer regions are still
optically thick and reprocess all the radiation that was previously
reprocessed by both the inner and the outer regions of the disk. Once also
the outer regions get deprived of their small grains, they too become
optically thin and the entire IR excess drops down. Note that the far-IR
excess for model F2 (with turbulence) is usually larger than that of model
F1. This is because even though the grains grow faster in model F2, they
also get swept up higher above the midplane so that the disk can reprocess 
a larger fraction of the stellar radiation.

The results of this section clearly show that the quick disappearance of the
small grains due to the efficient interaction between the differential
settling and the turbulent mixing is clearly in contradiction with
observations. The absence of turbulence may leave a minute population of
small grains, but is not very efficient in solving the problem fully.  In
the next section we discuss what we believe is the most likely solution
to this apparent contradiction with observations. In
Section \ref{sec-discussion} we discuss various other possible solutions.

\section{A simple local model including aggregate fragmentation}
\label{sec-destr}
So far we have taken a perfect sticking condition: if two grains/aggregates
collide, they stick and form a larger aggregate. However, for large
collisions velocities this is no longer true. In this section we show what
happens when aggregates are allowed to disintegrate upon collision.  This
replenishes the small grain size bins, and may go some way toward solving
the time-scale problem. Unfortunately the computational demand for such a
calculation is orders of magnitude higher than for a pure coagulation
calculation. Therefore we only show the results for a single vertical slice
(model SD1).

We include aggregate fragmentation in the very simplified way described in
Section \ref{subsec-destr}: if the collision energy exceeds a certain
limit, both grains are destroyed upon impact. We take as our model
parameters again the same parameters as in model S4 of
Section \ref{sec-models-slice}. 

\begin{figure}
\centerline{\includegraphics[width=9cm]{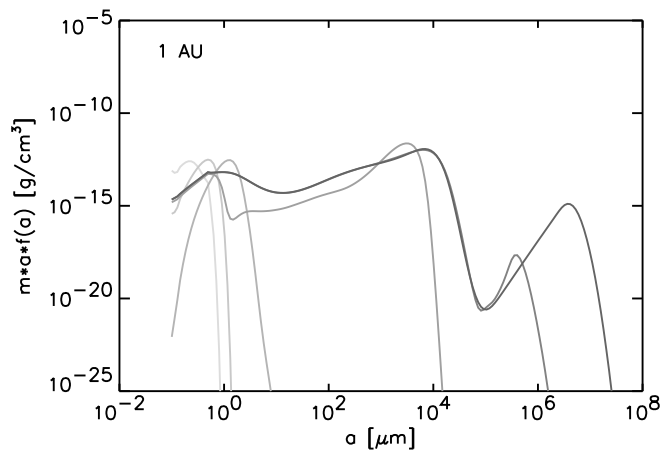}}
\caption{\label{fig-sd1}The time-evolution of the distribution function for
model SD1 (including all growth processes and aggregate fragmentation). This
slice has only been evolved up to $8\times 10^5$ years (darkest curve).}
\end{figure}

From Fig.~\ref{fig-sd1} we see that, while the initial stages of growth are
identical to those of model S4, very quickly the replenishment of small
grains due to fragmentation starts to take place. After about $10^4$ years a
semi-stationary state is reached for sizes below $a\lesssim 1\cm$. This is
an equilibrium between grain growth and grain fragmentation. As grain
aggregates reach sizes of about 1 cm, most of the aggregates are destroyed
and their mass returned to monomers, where it immediately starts to
coagulate again. For grain sizes larger than 1 cm the distribution function
continues to evolve, forming a powerlaw `tail' with a growing peak value.
This is a tail distribution of `lucky' grains which managed to avoid an
encounter with an equal size collision partner and therefore avoided
fragmentation. These grains sweep up some of the small grains from the
semi-stationary distribution below 1 cm, and therefore manage to grow toward
larger sizes. By this time other important effects, which we haven't taken
into account yet, will become dominant, such as radial drift and the
consequent run-away growth. Also some of our equations start to become
invalid for such large particles, in particular our formula for the gas 
drag in the Epstein regime.

\section{Discussion}
\label{sec-discussion}

\subsection{Other mechanisms to keep up the small particle population}

The basic fact that coagulation will lead to a reduced optical depth
in disks has been noted earlier.  \citet{1984Icar...60..553W} noted
that the reduction in opacity may lead to a termination of turbulent
gas motions if those motions are driven by convection.
\citet{1988A&A...195..183M} forced a steady-state solution for their
coagulation calculations by assuming that the disk is constantly
supplied with new gas containing new small particles.  They also
discuss that the addition of new particles may lead to alternating
phases of high and low optical depth, and consequently of convection
driven turbulence.  However, while turbulence driven by convection may
be present in disks, other drivers for turbulence may be important as
well.  \cdrev{The magneto-rotational instability \citep{balbushawley:1991}
seems to be an important candidate.  For the presence of this
  instability, low gas column densities are required in order to allow
  cosmic rays and X rays to penetrate and ionize the gas. High gas
  columns can create
so-called dead-zones \citep{gammie:1996} in the disk mid-plane where
ionizing radiation (such as X-rays and cosmic rays) cannot penetrate.}
Furthermore, the idea of a constant inflow of small dust grains onto
the disk can only be valid for the early evolutionary phases of a
disk. Observations show that circumstellar disks at an age of several
million years still can be strongly flaring
\citep{2004A&A...423..537L}.  By this time, the parental cloud has
been largely cleared away and infall of new material as well as
accretion onto the star has virtually ceased.  The exact limit on the
amount of infalling material which can be relevant for the disk
opacity is not easily estimated.  It will depend on the speed at which
vertical mixing and turbulence can remove these grains from the disk.
Clearly, for newly added material, the removal time scale will be
smaller since the low dust density slows down coagulation.
Pending more detailed calculations of these effects, it seems to us
that the infall of fresh gas and dust will not solve the discrepancy
between the speed and effectiveness of coagulation on the one hand,
and infrared observations of disks on the other hand.

\subsection{The effect of the large grain population on the small grains}
The choice of upper boundary to the grain size ($a_{\mathrm{max}}$) in the
above presented simulations has a strong influence on the results for the
case in which both differential settling and turbulent mixing are included
into the calculation. If $a_{\mathrm{max}}$ is taken too small (for instance
1 mm only), the grain growth is artificially stopped at that size. This
means that a comparatively large population of grains remains at a size
around $a_{\mathrm{max}}$ (while it should have grown further), providing a
comparatively large cross-section for the depletion of the smaller (1
$\mu$m) grains. Therefore, choosing $a_{\mathrm{max}}$ to be 1 mm instead of
10 m or more, strongly enhances the depletion of the small grains. In our
simulations we therefore had to choose $a_{\mathrm{max}}$ to be very large
(we took it 100 m). But this introduces yet another problem: our equations
are no longer accurate at those large sizes. Moreover, large bodies
tend to drift inward toward the star at a very high pace. This is not
included in the present model, because each vertical slice is assumed to be
independent of the other slices. Future investigations will have to deal
with this problem in a proper way. However, it is not likely that that would
solve the
problem of the quick depletion of small grains, since even for the
non-turbulent case (which represents the slowest growth) the depletion is
quite strong. 

\subsection{The inner regions of the disk}
Close to the central star (the inner regions of the disk) the
depletion of small grains goes the quickest, since the Kepler time
scale is the shortest there. The problem of the quick depletion of
small grains seems therefore to be most acute in those regions. But
can we be sure that small grains indeed exist so close to the star?
The answer seems to come from very recent observations with the mid
infrared interferometer MIDI on the Very Large Telescope. Using this
instrument, van Boekel et al.~(Nature, submitted) separated the
correlated 8-13 $\mu$m flux from the total 8-13 $\mu$m spectrum for 3
Herbig Ae/Be stars. In this way they were able to single out the
spectrum from the inner 2 AU region of the disk. Although the 10
$\mu$m silicate feature in this correlated flux spectrum clearly
showed evidence for grain growth up to 2 $\mu$m, it also clearly
showed that grains of approximately 2 $\mu$m were still present and
their abundance is strong enough to produce a clearly discernible 10
$\mu$m feature. This is in clear contradiction with the
pure-coagulation models presented here, and reinforce our conclusion
that aggregate fragmentation (or some other resupply of small grains)
should play a major role in disks.

\subsection{Interpretation of the results}
The models of grain growth presented in this paper show that coagulation
happens on a time scale that is two to three orders of magnitude too short
to be consistent with observations of T Tauri star disks or the disks around
Herbig Ae/Be stars. Turbulent mixing combined with coagulation through
differential settling is highly efficient at removing the small grains from
the disk at all heights above the midplane. One may argue that some disks
may have regions of zero turbulence (for instance the `dead zone' introduced
by Gammie \citeyear{gammie:1996}), which may decrease the efficiency of the
grain growth. But even if we only have differential settling, but no
vertical mixing (as in model S2) then only a tiny population of small grains
remains in the disk, containing less than $10^{-6}$ of the original
population of small grains. The full disk model F1, which is 
without turbulence, clearly shows that at typical ages of T Tauri stars
(around 1 Myr), almost no IR emission is left, and for the model F2 (with
turbulence) this is even more dramatic. If anything, the SED looks
like that of a debris disk instead of a T Tauri star disk. But even if
we compare the SED at $10^4$ years to observed SEDs, we find that the
dip in the SED at near- to mid-IR wavelength is rather untypical for
most T Tauri stars. 

There are, however, a few examples of objects which show conspicuous
near/mid-IR dips. A strong near/mid-IR dip in the SED of TW Hydra was
interpreted by Calvet et al.~(\citeyear{calvetdalessio:2002}) as a
signature of a planet which has cleared out the inner disk. The
synthetic SED of model F2 at $10^4$ years, however, shows a similar
dip, and therefore dust coagulation could be an alternative
explanation. HD100546 is a Herbig star which also features a
conspicuously weak near-IR excess, which was attributed to a huge gap
in the disk (Bouwman et al.~\citeyear{bouwmandekoter:2003}).  Here
again, dust coagulation may be an alternative explanation. 

From the results presented in this paper it seems unavoidable that some form
of replenishment of small grains is needed to make the model calculations
comply with the observations. The only other possibility is that the
sticking probability is enormously reduced by some process. Since we are not
aware of a process capable of reducing the sticking probability by such a
dramatic amount, we believe that replenishment is the only
solution. Replenishment by destructive collisions seems to be the most
natural way to prevent the small grains from disappearing entirely.  In this
paper we demonstrated that this could work if we assume very low binding
energies of the grains. The process of cratering (a small particle impacting
on a bigger one and creating a certain amount of impact debris) may be a
better way to produce small grains, but we defer a more detailed
implementation of aggregate fragmentation to a future paper.

\section{Conclusions}
In this paper we have modeled dust coagulation in protoplanetary disks and
computed the SEDs of such disks. We qualitatively compare these SEDs
to what is typically observed from T Tauri star and Herbig Ae/Be star disks,
and we conclude that if coagulation is allowed to proceed
unhindered (i.e.~without fragmentation of aggregates), then the small
grains are depleted on a time scale that is three orders of magnitude
too short to be consistent with these observations.
We have included three coagulation mechanisms
in this model (Brownian motion, differential settling and turbulence).
The inclusion of only the first two, well understood, processes
already shows that the strong and rapid depletion of small grains is
unavoidable unless small grains are somehow replenished.  The inclusion
of a little bit of turbulent mixing will only aggravate matters. 
It is very difficult to slow this process down, even by
grain charging or other mechanisms. Either the grain sticking
efficiency is many orders of magnitude less than currently assumed, or
the small grain population must be replenished in some way.  We suggest
that aggregate fragmentation could be such a mechanism.  We present a
highly simplified model for this and show that a semi-stationary
equilibrium sets in in which coagulation and fragmentation are balanced
for an extended amount of time.  Whether this is the solution to the
paradox remains unclear and requires much more detailed simulations.

\begin{acknowledgements}
We wish to thank the referee S. Weidenschilling for a fast and
insightful report which helped to improve the paper.
\end{acknowledgements}

\appendix

\section{Turbulence-driven coagulation}\label{app-voelk}

Turbulent motions in the gas can cause collisions between dust
particles.  The basic reason for this is that particles with different
$\siggas/m$ ratio couple to eddies of different size and therefore
acquire random velocities.  To calculate the motions caused by
turbulence, it is necessary to integrate over the contributions of all
eddies from the largest scales down to the smallest scales which are
set by the condition that the turbulent Renolds number Re equals 1.

\def\vturb{\ensuremath{v_\mathrm{edd}}}
\def\tauturb{\ensuremath{t_\mathrm{edd}}}
\def\vturbs{\ensuremath{v_\mathrm{edd}^\mathrm{s}}}
\def\tauturbs{\ensuremath{t_\mathrm{edd}^\mathrm{s}}}
\def\vturbnull{\ensuremath{v_\mathrm{edd}^\mathrm{0}}}
\def\lambdaturbnull{\ensuremath{\lambda_\mathrm{edd}^\mathrm{0}}}
\def\tauturbnull{\ensuremath{t_\mathrm{edd}^\mathrm{0}}}
\def\tfric{\ensuremath{\t_\mathrm{fric}^\mathrm{0}}}

Calculating the gas-dust interaction is involved and has been covered
by previous authors \citep[e.g.][]{weidenschilling:1977,cuzzidobrchamp:1993}.
Here we only describe the basic recipe implemented in our code.

The main particle property that enters the calculation is the stopping
time of of a particle which is given by
\begin{equation}
\label{eq:55555}
t_i=\frac{3}{4}\frac{m}{\siggas}\frac{1}{\rhogas\cs}
\end{equation}
which physically is the time in which a particle reacts to changes in
the motion of the surrounding gas.  Specifically, for a given
turbulent eddy, this time scale indicates if the particle will follow
the eddy motion or not.

In order to derive the average relative velocities resulting from this
mechanism, the full structure and spectrum of the turbulence must be
known.  Turbulence is often characterized by the velocity and time
scale of the largest eddies \citep[see][]{duldomsett:2004}
\begin{align}
\label{eq:1}
\vturbnull &= \alphaturb^q \cs\\
\tauturbnull &= \frac{2\pi}{\Omegak}
\end{align}
where $q$ is a turbulence parameter between 0 and 1.  Following
Schraepler and Henning (\citeyear{schreaphenn:2004}) we take it to be $q=1/2$.  The
energy then cascades down to small sizes, until the flow becomes
laminar at a turbulent Reynolds number Re$=1$ in a gas with viscosity
\begin{align}
\label{eq:2}
\vturbs &= \vturbnull \mathrm{Re}_0^{-1/4}\\
\tauturbs &= \tauturbnull \mathrm{Re}_0^{-1/2}
\end{align}
with
\begin{align}
\label{eq:3}
\mathrm{Re}_0 = \frac{\rho \vturbnull \lambdaturbnull}{\eta} \approx
\frac{\rho (\vturbnull)^2\tauturbnull}{\eta} \quad .
\end{align}

The random motions of the particles must then be calculated by
integrating over the contributions of the different eddy scales.  For
a Kolmogorov spectrum of the turbulence, this has been done
numerically by \citet{voelkmorroejon:1980} and
\citet{1988A&A...195..183M}.  \citet{1984Icar...60..553W} has fitted
the numerical results with a simple analytical formula, which has also
been used for example by \citet{2001ApJ...551..461S}.  We will also
adopt it, but with two modifications:
\begin{enumerate}
\item\label{item:1} Because of the lower cutoff to the eddy size
  spectrum, no random velocities relative to the gas are introduced by
  turbulence for particles with a stopping time below the turnover
  time of the smallest eddy, \tauturbs.  This has been noted by
  \citet{1984Icar...60..553W} and discussed in detail by
  \citet{1988A&A...195..183M}.  We use the
  limiting case provided by the latter authors.
\item The analytical fit produces an overshoot in the limit
  $t_2\to\tauturbnull$.  If $t_1\ll t_2$, the analytical fit given by
  \citet{1984Icar...60..553W} and also used by
  \citet{2001ApJ...551..461S} leads to $\vturbnull=3\vturbnull$ while
  the numerical results by \citet{voelkmorroejon:1980} only exceed
  $\vturbnull$ by a few percent.  In order to avoid effects cause by
  this incorrect high collision speeds, we therefore limit the
  relative velocity caused by turbulence to \vturbnull.
\end{enumerate}

With both modifications, the recipe for the turbulent collision
velocities becomes
\begin{equation}
\label{eq:4}
\Delta \vturb
\left\{ 
\begin{array}{l@{\quad:\quad}l}
\frac{\vturbs}{\tauturbs} |t_1-t_2|
\sqrt{\frac{\ln \mathrm{Re}}{2 \sqrt{\mathrm{Re}}}
\frac{\tauturbnull}{t_1+t_2}}
& \mbox{if $t_1,t_2\le\tauturbs$} \\
\vturbnull
& \mbox{if $t_1\le\tauturbnull\le t_2$} \\
\vturbnull \frac{\tauturbnull(t_1+t_2)}{2t_1t_2}
& \mbox{if $\tauturbnull\le t_1,t_2$} \\
\vturbnull \min \left\{1,\frac{3t_2}{t_1+t_2}\sqrt{\frac{t_2}{\tauturbnull}}\right\}
& \mbox{otherwise}
\end{array}\right.
\end{equation}

\section{Numerically solving the coagulation equation}\label{app-numerics}
The numerical solution of Eq.~(\ref{eq-smoluchovski}) is a subtle matter.
Consider a discretized grain mass grid $m_i$ with $i\,\in [1,N]$.  The first
integral on the right hand side can be converted into a sum over $m'=m_k$
(for fixed $m=m_i$) with $k=1\ldots l$ where $l$ is the highest index for
which $m_l\le m_i$. Unfortunately the value of $m_i-m_k$ lies generally not
exactly at a discrete mass point 
and therefore the value of $f(m-m')\equiv f(m_i-m_k)$ must be obtained by
interpolation.  Since $f$ can vary extremely strongly, the interpolation is
best done in $\log(f)$ instead of in $f$. Numerical practice has shown that
two-point interpolation makes the algorithm more stable than four-point
interpolation. The last point in the integral (i.e.~the numerical sum) is
located at $m/2$, and in that case $m'=m-m'$, i.e.~both $m'$ and $m-m'$ are
located in between two mass points and similarly an interpolation needs to
be done. The integral in the second term on the right hand side of
Eq.~(\ref{eq-smoluchovski}) is easier, since no interpolation is needed.

\subsection{Renormalization}
The right-hand-side of Eq.~(\ref{eq-smoluchovski}) consists of a gain and a
loss term. The gain term describes how much matter enters a certain mass bin
through coagulation of smaller particles, while the loss term describes how
much matter leaves the mass bin through coagulation of this matter with
particles of any other size. Typically these two terms are very large
numbers which cancel each other almost entirely, except for a tiny amount.
It is this tiny amount that is the crucial source term for the coagulation
equation. This near cancellation happens then when the gain of matter in a
bin is dominated by coagulation between large and small particles. The
reason for this near cancellation is best described with an example.
If a rock of 1 kg that hits a dust particle of 1 micron size, formally
the rock increases in mass (albeit by an extremely small amount). The rock
is therefore removed from its mass bin and put into the mass bin a few
picogram toward larger mass. Since the 1 kg rock may 
collide with trillions of 1
micron size particles, the gain and loss terms in the kg mass bin are
huge, but virtually identical. The minuscule difference between these
two terms determines the eventual growth from 1 kg to 2 kg after the rock
collects 1 kg worth of micron size particles. 

Numerically this poses a significant challenge. If one simply computes
the gain and loss terms, the near cancellation goes astray once the
cancellation happens beyond the 14th digit. In effect the near
cancellation turns into a perfect cancellation, which is incorrect. To
solve the problem the integrals of the gain and loss terms have to be
calculated simultaneously. The integrands of both integrals are
calculated at the same time, and then subtracted and collected into a
single integral. At each $m'$ it is checked if the two terms tend to
produce a near cancellation. If so, their difference is recomputed
using a renormalization technique:
\begin{equation}
\begin{split}
f(m')&f(m-m')\sigcoag(m',m-m')\Delta v(m',m-m') -  \\
f(m')&f(m)\sigcoag(m',m)\Delta v(m',m) dm' \\
\simeq &  -f(m')(m-m') \\
& \frac{d(f(m'')\sigcoag(m',m'')\Delta v(m',m''))}{dm''}
\comma
\end{split}
\end{equation}
which uses l'Hopital's rule. This renormalized version of the integrand is
valid only when a near-cancellation takes place, and remains numerically
well-determined even for extreme cases of near-cancellation.

\subsection{Mass conservation}
By performing the integration in the above described way, total dust mass is
not necessarily perfectly conserved. Small errors can increase or decrease
the total mass, and since these errors are generally systematic, the risk is
high that the dust mass will unlimitedly grow or diminish as the simulation
progresses. In our code this is avoided by making small corrections. Denote
$S(m)$ as the right-hand-side of Eq.~(\ref{eq-smoluchovski}). Define now
the following two functions:
\begin{eqnarray}
S_{+}(m) & = & \left\{\begin{matrix}
S(m) & \hbox{where} & S(m)>0 \\
0 & \hbox{where} & S(m)\le 0 \\
\end{matrix}\right.\\
S_{-}(m) & = & \left\{\begin{matrix}
0 & \hbox{where} & S(m)>0 \\
S(m) & \hbox{where} & S(m)\le 0 \\
\end{matrix}\right.
\end{eqnarray}
The new right-hand-side $S_{\mathrm{new}}(m)$ is now:
\begin{equation}
S_{\mathrm{new}}(m) = S_{-}(m) + \chi S_{+}(m) 
\end{equation}
where $\chi$ is defined as:
\begin{equation}
\chi = -\frac{\int m\,S_{-}(m)\,dm}{\int m\,S_{+}(m)\,dm}
\end{equation}
This is generally a tiny correction, and it guarantees mass conservation
and thereby prevents unphysical build-up or loss of matter. 

\subsection{Time step determination for the coagulation}
Since the evolution of the grain size distribution proceeds each time step
by adding a source term with the results of the integrals, the magnitude of
this time step is set by:
\begin{equation}
\Delta t = \xi\; \mathrm{min}\left(\frac{f(m)}{|S(m)|}\right)
\comma
\end{equation}
where $0<\xi\le 1$ is an accuracy parameter which we usually set to $0.3$ in
our simulations. Typically this time step is shortest at the smallest $m$,
so that the evolution at small $m$ limits the time step of the entire
simulation. This could in principle be prohibitively short compared to the
total time we wish to evolve our simulation ($10^7$ years).  Fortunately, as
the aggregates grow, the smallest mass bins are quickly depleted and can be
set to zero once the value of $f(m)$ drops below some floor value. The time
step then does not need to be limited anymore by these small mass bins. In
practice this means that as the time progresses, the time step becomes
larger and $10^7$ years can be reached without many problems.

Sometimes, the simulation may get temporarily stuck at relatively small time
steps, despite the above mentioned time stepping method. This is caused by the operator
splitting between the settling/mixing and the coagulation. While the
coagulation equation may attempt to empty a certain mass bin, the settling
and mixing may fill it up again. This typically happens at the midplane,
where settling and vertical mixing may replenish the mass in a certain mass
bin by transporting it down to the midplane from higher altitudes. Since the
settling and mixing are numerically simulated in an implicit way, allowing
the time step to exceed the Courant condition for for settling and mixing by
many orders of magnitude, this can cause the instant refilling of the mass
bin after the coagulation equation has tried to deplete it. In practice,
however, the simulation never gets entirely stuck, and at some point manages
again to increase the time step to large enough values to allow the
simulation to end in about 20 minutes per vertical slice on a Pentium 4
processor. Therefore it is manageable to simulate an entire disk consisting
of 20 vertical slices in about 8 hours CPU time.

However, once the aggregate fragmentation is included, the small grains remain
present, and the time step remains limited by the smallest mass bin. The
above time stepping method then does not work anymore, and the simulation may
take large amounts of CPU time even for a single vertical slice. A fully
implicit treatment of the coagulation/fragmentation may then be necessary to
prevent excessive computational costs.

\section{Test cases}\label{app-tests}
It is not straightforward to test the numerical algorithm described here
with the physics we include into the coagulation equation, since no analytic
solutions exist for the kernel we use. Lai et al.~(\citeyear{laifried:1972})
have presented analytical solutions to the coagulation equation with a
Brownian motion kernel, but these solutions are only valid in limiting cases
and they contain undetermined constants.

For simplified kernels, various analytic solutions to the coagulation
equation exist. The two most well-known cases are the case of
$K(m_1,m_2)=(m_1+m_2)\,A$ (Safronov \citeyear{safronov-book}; henceforth
test 1) and the case of $K(m_1,m_2)=A$ (Smoluchowski
\citeyear{smoluchowski:1916}; henceforth test 2). In both cases $A$ is an
arbitrary constant which we take to be $A=1$. Both tests were extensively
discussed by Ohtsuki et al.~(\citeyear{ohtsuki:1990}). They find that in
particular for test 1 numerical algorithms generally deviate significantly
from the analytic solution if the coordinate spacing in mass is too coarse
($m_{i+1}/m_{i}\gtrsim \sqrt{2}$). This seems to be a particularly tough test
case. In contrast they find that test 2 is much less challenging for
numerical algorithms.

\begin{figure}
\centerline{\includegraphics[width=9cm]{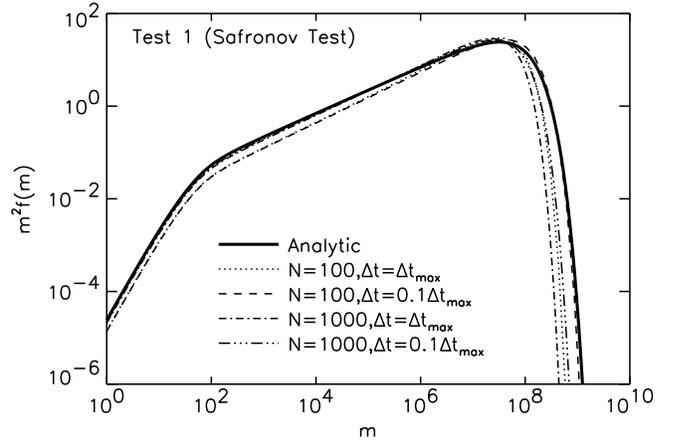}}
\caption{\label{fig-test-1}A snapshot of the distribution function of test 1
at dimensionless time $\tau=9$, for various values of the grid resolution
and time step. The symbol $\Delta t$ is the time step ($\Delta
t_{\mathrm{max}}\equiv\mathrm{max}(f(m)/S(m))$) and $N$ is the number of grid
points in $m$ ($N=100$ corresponds to $m_{i+1}/m_{i}=\sqrt{2}$, i.e.~the
highest resolution used by Ohtsuki; $N=1000$ corresponds to
$m_{i+1}/m_{i}=1.03517$). The solid line is the analytic solution of
Safronov.}
\end{figure}

We have done both tests with our algorithm, following Ohtsuki's test
procedure and using the analytic solutions presented in that paper. We find
similar results as they find. For test 1 we get qualitative agreement, but
the precise location of the peak of the distribution seems to depend on mass
grid resolution and time step size. In Fig.~\ref{fig-test-1} the results are
shown at a single given time, for four simulations: for 100 and 1000 grid
points and for the largest possible time step ($\Delta
t_{\mathrm{max}}=0.3\,\mathrm{max}(f(m)/S(m))$, where the factor $0.3$ is
our ``safety parameter'') and for 0.1 time that value. The modest but
non-negligible deviations found in test 1 are slightly troubling. But since
Ohtsuki et al.~experience similar problems and the kernel of test 1 is very
unlike the kernel we use in our models, we are reasonably satisfied with the
results of these tests.

\begin{figure}
\centerline{\includegraphics[width=9cm]{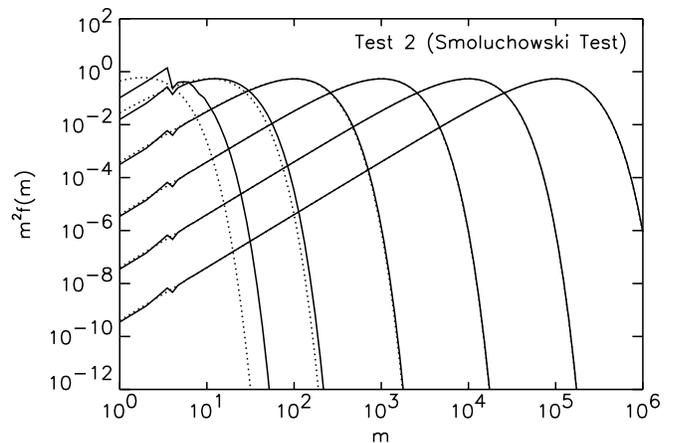}}
\caption{\label{fig-test-2}The time evolution of the distribution function
in test 2 for the model (solid line) compared to the analytic solution of
Smoluchowski (dotted line). The snapshots are at dimensionless time
$\tau=1,10,10^2,10^3,10^4,10^5$ from left to right. Due to different initial
conditions the agreement is not good at $\tau=1$ but becomes better as time
progresses because the initial conditions are `forgotten' by the system. The
tiny wiggle in the solid curve around $m=4$ is a remainder of the initial
conditions that apparently the algorithm does not `forget'.}
\end{figure}

For test 2 we get excellent agreement for moderate grid resolution (100 grid
points) and for the largest possible time step size described above. The
results are shown in Fig.~\ref{fig-test-2}. Initially the difference between
our model and the analytic solution is relatively large, but this is because
our coagulation equation is formulated in a continuous form while the
Smoluchowski solution is based on the discrete form of the equation. We
therefore start with a different initial condition as in the analytic
solution of Smoluchowski and logically we get initially somewhat different
results. But as time progresses, the initial conditions are `forgotten', and
the model result and the analytic solution start to agree better and better,
ending with almost perfect agreement in our last time step.

These positive test results for simple kernels give hope that the full code,
with dust settling, mixing and coagulation through Brownian motion and
differential settling, also works well. Moreover, the stability and
reliability of the code was confirmed by reproducing the results presented
in this paper with different grid resolutions (both in $z$ and in $m$), and
by using different time steps. But of course the only way to be sure that
the code is working properly is a comparison of our code with various
independent codes written by independent authors. Such a comparison has not
been carried out. There is, however, an interesting way to test the code
independently: by comparing the average size of grains rained out onto the
midplane (in the absence of turbulence) to the result of the one-particle
model. From the one-particle model (Section \ref{sec-safronov}) it follows
that the size of the grain, once it reaches the midplane, is
$a=0.01\,\Sigma/8\rho_{d}$ for compact grains. It should be kept in mind
that this result was obtained by assuming that only the test particle rains
down, while the other particles remain suspended in the disk. We can
simulate this effect in the full code by taking two populations of grains: a
fixed population (equal to the initial grain population) and an evolving
population. The evolving population only grows by colliding with the fixed
population, while the fixed population is not changing during the
simulation. This does not conserve mass, but it does simulate the
conditions similar to the one-particle model. We carried out this test for
$\Sigma_{\mathrm{gas}}=100$ g/cm$^2$ at $R=1\AU$ with $\rho_{d}=3.6$, and we
find that the model produces a sharp peak distribution near $a=0.35$mm,
which is virtually identical to the analytic result ($a=0.3472$mm).
That it is slightly larger is due the fact that the smallest grains
still have a finite geometrical cross section.

\end{document}